# Energetics of Protein-DNA Interactions

Jason E Donald*
Department of Chemistry and Chemical Biology, Harvard University, 12 Oxford St.
Cambridge, MA 02138

William W Chen
Department of Biophysics, Harvard University
Cambridge, MA 02138

Eugene I Shakhnovich
Department of Chemistry and Chemical Biology, Harvard University, 12 Oxford St.
Cambridge, MA 02138

*To whom correspondence should be addressed:
jdonald@fas.harvard.edu

**Abstract**

Protein-DNA interactions are vital for many processes in living cells, especially transcriptional regulation and DNA modification. To further our understanding of these important processes on the microscopic level, it is necessary that theoretical models describe the macromolecular interaction energetics accurately. While several methods have been proposed, there has not been a careful comparison of how well the different methods are able to predict biologically important quantities such as the correct DNA binding sequence, total binding free energy, and free energy changes caused by DNA mutation. In addition to carrying out the comparison, we present two important theoretical models developed initially in protein folding that have not yet been tried on protein-DNA interactions. In the process, we find that the results of these knowledge-based potentials show a strong dependence on the interaction distance and the derivation method. Finally, we present a knowledge-based potential that gives comparable or superior results to the best of the other methods, including the molecular mechanics force field AMBER99.

# Introduction

Specific protein-DNA interactions are necessary for the proper function of the cellular machinery. Gene regulation depends on transcription factors being able to find particular DNA sequences, and a restriction enzyme must only cut the correct sequence. To have a better microscopic understanding of these interactions, reliable theoretical models are needed to describe the energetics of protein-DNA binding. For example, such a model could be used to better understand how DNA-binding proteins select the correct binding site out of a huge number of incorrect sites. Several different structure-based models of protein-DNA interactions have been proposed, but it is not clear how accurately they describe protein-DNA interactions nor has a systematic and rigorous method been proposed to compare these models.

In addition to gaining a better understanding of these important interactions, the unique features of protein-DNA interactions provide additional motivation for investigation. These unique features are most apparent when one compares the study of protein-DNA interactions to two other fields where molecular interactions play an important role: protein folding (self-interactions along a polymer chain) and protein-protein binding (interactions between two macromolecules of the same type). In protein fold prediction, the goal is to discriminate the native structure from an enormous number of very different misfolded structures. For protein-protein binding, a model needs to predict not only which proteins interact but also find the correct orientation of the interaction out of the large number of possible misinteracting structures. Because of the linearity and limited alphabet of DNA, protein-DNA interactions are much simpler to study. In the readout of a DNA sequence by a protein, the model is expected simply to predict the DNA sequence to which the protein will preferentially bind. For a DNA sequence of length N, where N is the number of bases in contact with the protein (often 10-20), there are only $4^N$ possible sequences. If one considers a protein interacting with a single genome, the problem is further restricted from $4^N$ sequences to the number that is biologically relevant. This makes protein-DNA interactions an excellent system to compare the different theoretical models, and the comparison may provide useful information for the study of protein folding and protein-protein interactions.

The problem of protein-DNA specificity can roughly be framed by dividing the interaction into two energetic components: direct and indirect readout. When a protein is bound to DNA, the DNA is often bent away from the lower energy unbound conformation. DNA sequences pay different energetic penalties to be deformed in the bound state because of their different rigidities and conformational preferences. DNA-binding proteins can use these different penalties to prefer specific sequences. This preference is referred to as indirect readout.

The second component, direct readout, refers to the specific energetic interactions between the protein and DNA. These include hydrogen bonding, electrostatic, and hydrophobic interactions. As an example, hydrogen bonding allows arginine amino acids to generally prefer guanine bases over cytosine.

Many existing energy models either use separate direct and indirect readout terms(1,2) or ignore indirect readout altogether(3-7). Those methods that do consider indirect readout have used two different types of methods: the knowledge-based method of Olson, et al.(8) or a molecular mechanics method(9-11). An initial comparison between these two methods of indirect readout has recently been carried out(12). As there has not yet been a systematic comparison of direct readout, here we will focus on this very important component of protein-DNA interactions.

Direct readout has been studied using the same two types of methods that were used for indirect readout, molecular mechanics and knowledge-based methods. We will consider the results of both types of methods. Quantum mechanical calculations(13) have also been used, but these are currently too computationally intensive to represent protein-DNA interactions in an aqueous environment.

Molecular mechanics-type potentials can be divided into two subgroups based on the method use to fit the parameters. Standard molecular mechanics potentials use experimental and theoretical results from small molecules to train parameters for macromolecular interactions. The ROSETTA method(1,3), on the other hand, also is trained using experimental results from macromolecules. Both assume that experimental measures from one context will accurately predict measures in another context.

The other major type of methods, knowledge-based potentials, uses the frequency of contacts between different residues or atoms in known crystal structures to predict the interaction energy. If residues or atoms are often in contact, one expects that the energy of interaction between them is favorable. If they are rarely found in contact, one expects that the energy is unfavorable. These methods assume that despite the chemical connections of the polymer chains, the number of contacts in the database will well represent the interaction energies. Three different knowledge-based potentials have been used(4-6,14), but there are other types of knowledge-based potentials that, while successfully used in protein folding(15,16), have yet to be used to study protein-DNA interactions.

We present two of these previously unstudied knowledge-based potentials and compare them to existing knowledge-based and molecular mechanics-type potentials. We show that knowledge-based potentials compare well to other potentials but that the way in which a knowledge-based potential is derived and its treatment of the interaction distance are vital to its performance.

# Methods
## I. Protein-DNA energetic models
### A. Knowledge-based potentials.

When developing a knowledge-based potential, there are several details of the model that need to be chosen. We begin with what is termed the quasichemical method. We then test variations of this method that modify the potential derivation equation or the molecular representation. The first three methods all consider interactions between non-hydrogen ("heavy") atoms while the last method considers interactions between protein residues and DNA bases. For each method we also systematically modify the interaction distance cutoff. To our knowledge, neither the heavy-atom quasichemical potential nor the μ potential, a different knowledge-based potential derivation, has been used previously for protein-DNA interactions.

### i. Different potential derivations

The first method, the quasichemical method(16,17), expects that an interaction with zero energy would be seen in contact as often in known structures as it would be in structures where the contacts were randomly shuffled between the atoms. When there are more than the expected number contacts between atom pairs in known structures, the interaction is predicted to be attractive. Conversely, fewer contacts than the expected number result in a predicted repulsive energy. The method calculates the interaction energy ε between protein atom type i and DNA atom type j by:

$$\varepsilon(i,j,d) = -RT \ln \frac{N(i,j,d)}{N(d)\chi_i \chi_j} \quad (1)$$

where d is the distance bin in which the contact occurs, R is the gas constant, T is the temperature, N(i,j,d) is the number of contacts of this type occurring in the training set in the distance bin, N(d) is the total number of contacts in the distance bin, and $\chi_m$ is the fraction of atoms that are of type m in either protein or DNA. For convenience, we use reduced units where RT is unity.

The denominator of the log function expresses the expected number of contacts between the two atom types in the particular distance bin and is known as the reference state. The quasichemical method's reference state assumes a random shuffling of the protein atoms and the DNA atoms.

A second method known as the DFIRE potential(14,18) calculates the interaction energy using a different reference state:

$$\varepsilon(i,j,d) = -RT \ln \frac{N(i,j,d)}{N(i,j,d_{cut})\left(\frac{r(d)}{r(d_{cut})}\right)^{\alpha}\left(\frac{\Delta r(d)}{\Delta r(d_{cut})}\right)} \quad (2)$$

where $d_{cut}$ is the reference distance bin, r(d) is the distance of the midpoint of the contact bin, and Δr(d) is the width of the contact bin. DFIRE's reference state assumes that interactions are short range such that at a sufficiently distant $d_{cut}$ the atoms will be found in contact as if there were no interaction potential between them. The number of contacts is then normalized by the volume in the different bins. The adjustable parameter α is chosen to be 1.61, reflecting the finite-size effect of confining the atoms within protein-sized spheres (14,18,19).

We have found a third method, known as the μ potential, to be useful for modeling protein folding(15,20,21). It is a generalization of the topological Gō potential used to make the native state the minimally frustrated global minimum. The interaction energy in this potential is:

$$\varepsilon(i,j,d) = \frac{-\mu(d)N(i,j,d) + (1-\mu(d))N^*(i,j,d)}{\mu(d)N(i,j,d) + (1-\mu(d))N^*(i,j,d)} \quad (3)$$

where $N^*(i,j,d)$ is the number of non-contacts. $N^*(i,j,d)$ is equal to the total number of possible i-j atom pairs in the complex minus $N(i,j,d)$, the number of i-j pairs in contact in distance bin d.. The parameter μ is chosen to make the mean ε(i,j,d) zero for each distance bin.

The fourth potential, the residue potential, uses a different molecular representation. We modify the quasichemical method to consider residues and bases instead of protein and DNA heavy atoms. The distance between a residue and a base is taken to be that between the protein $C_\beta$ atom ($C_\alpha$ for glycine) and the sugar contacting atom of the base (N9 for A,G and N1 for C,T).

As a final control, we also present the "non-specific" potential. It simply gives an energetic value of -1 for each protein-DNA contact within a certain distance cutoff. The non-specific potential is useful for distinguishing between features of a potential that give it true discriminatory power and features that can be trivially attributed to contact density or contact number.

**ii. Common derivation details**

The training and testing sets include 163 x-ray crystallographic protein-DNA structures taken from the PDB(22). If a structure contains a protein sequence with a BLAST E-value match to either the training or testing set of less than $10^{-10}$ and a DNA sequence with a BLAST E-value match less than $10^{-5}$, it was excluded from the training set. For heavy atom potentials (quasichemical, DFIRE, and μ potentials) the atom typing scheme for proteins is taken from previous work(20) adding four heavy atom types for the protein backbone. The DNA atom types were chosen for this work (Supplemental Figure 1).

**iii. Distance treatments**

To select the distance bins for a knowledge-based potential, we try two different distance binning methods. The first distance binning method, used for the quasichemical, residue, and nonspecific potentials, has a single distance bin. The bin begins at 0 Å and the maximum distance cutoff is increased in steps of 0.1 Å from 3 to 15 Å. The residue and nonspecific potentials use this treatment exclusively because of their simpler representations. For the quasichemical, DFIRE, and μ potentials we use a multiple bin model. The first bin was chosen to be 0-3 Å to provide sufficient contact statistics. The remaining bins are in 0.5 Å steps (3.0-3.5 Å, 3.5-4.0 Å, etc.) up to 15 Å. Other step sizes were considered, but 0.5 Å appears to provide an acceptable compromise between having detailed distance information and limited statistics. Because the original DFIRE potential uses a bin from 14-15 Å for $d_{cut}$ we also use a 1 Å final bin for the DFIRE potential.

**B. Molecular mechanics-type potentials**

We also consider the results of the AMBER99 potential(23) as implemented in TINKER(24). While we did not explicitly consider hydrogens in the knowledge-based potentials, they are necessary for molecular mechanics calculations. Many crystal structures, however, do not include hydrogen atom positions. To provide the locations of hydrogen atoms, the atoms were placed by TINKER and minimized to an RMSD of 0.01 using the NEWTON function(25). Hydrogen position minimization is carried out in the gas phase. As explicit waters and ions are not modeled in the knowledge-based potentials and often not available in crystal structures, these small molecules are removed from the AMBER calculations.

For each structure, two AMBER99 energy calculations are then carried out. First, the energy is recorded for the initial structure (with minimized hydrogens) using the GB/SA solvation model(26). Next, the full structure is minimized using the same solvation model. Minimization is limited to 100 steps. This number of steps allows the structures to approach or reach a local minimum in a reasonable amount of time. The energy of the minimized structure is then recorded.

**III. Tests of the potentials**

We present three metrics to compare the accuracy of a given protein-DNA potential. The first is a specificity test: does the potential predict the crystal structure DNA as the lowest energy sequence? A potential is asked to rank the crystal structure DNA below all others, but it is not necessarily required to reproduce the energetics in a quantitative fashion. The second and third metrics assess the correlation between the predicted energy and experimental measurements. Because experimental free energies of binding are available for certain sequences and binding proteins, this metric does demand that the potential accurately reproduce the energetics, at least for the tested protein-DNA pairs. We will show, however, that scoring high on one of the latter two metrics, while necessary, is by no means a sufficient condition to ensure that the potential is truly reflecting physical energies.

In order to compare direct readout energy functions alone without considering the different indirect readout models, we only calculate the energies of rigid structures and their computationally mutated complements. The only exception is that structural minimization is used when testing AMBER. Relaxation of the structures would require an accurate and correctly weighted indirect readout energy function. In addition, others have found that using current potentials, relaxation of crystal structures actually decreases the predictive value of these potentials(1,9). Therefore, to change the structure to represent a DNA mutation, we simply replace the crystallized DNA base pair with the new base pair. To replace a base, the original base nitrogen atom bonded to the sugar and the two base carbon atoms bonded to this nitrogen are aligned with the corresponding atoms of the new base. This preserves the sugar base bond and the base planar angle. A representative substitution of base pairs is shown in Figure 1. Standard base structures used for the replacements were taken from 3DNA(27).

**A. Prediction of protein-DNA specificity**

To represent the accuracy of the lowest energy DNA sequence prediction, we count the number of mismatches between the predicted sequence and the sequence found in the crystal structure. This metric is known as the Hamming distance. For each structure, we only consider sequence positions that are in contact with a protein. A position is labeled as "in contact" when there is at least one protein atom within 5 Å of the base pair in the crystal structure. Because we compare sequences from systems where different DNA sequence lengths are bound, we normalize the Hamming distance by the number of bases in contact. We then average these distances to give the final score. A random selection of bases would be incorrect three out of four times giving a normalized Hamming distance of 0.75.

The training set includes all structures in the data set (Supplemental Materials) except the nine structures in the testing set (1a02, 1a3q, 1ckq, 1ecr, 1lmb, 1run, 6cro). The testing set was selected from a structural classification of DNA-binding proteins(28). Because it is unclear what the maximum distance cutoff should be, for each potential, the protein-DNA specificity test was used on the training set to select an optimal distance parameter. We select a single distance cutoff for each method to allow a straightforward comparison with other methods.

In order to determine the Hamming distance for a given potential and structure, all DNA sequences need to be considered. Normally, this would involve $4^N$ different protein-DNA energy calculations, where N is the length of the DNA sequence. The knowledge-based potentials, however, are pairwise and only consider interactions between the protein and DNA. In these energy models, the replacement of a single base-pair at a given position solely affects the interaction energy between the protein and that position. We therefore consider the three possible single mutations at each position. To calculate the final energy, we simply add the change in energy of each single mutation from that of the crystal structure. This requires a total of only $3N+1$ energy calculations.

For calculation of energies using AMBER, hydrogen positions were always minimized first. This is due to the special distance-dependent nature of AMBER, as small changes in distances between atoms can impact energetics greatly. The implementation of minimization procedures means that a mutation at one base possibly impacts the energy at another base; in other words, the results of mutations at different base positions will be non-additive. Most often, however, a given protein side-chain interacts directly with only a single base pair. This should mean that a base mutation will not strongly affect the direct readout protein-DNA interaction energy of another base mutation. For this test, we assume that any such non-additivity of mutations will be small. Others have found this to be a good approximation for a different type of molecular-mechanics type potential(1). As in the case of the knowledge-based potentials, we calculate $3N+1$ energies and add the predicted energy changes of the base mutations from the crystal structure.

**B. Prediction of total free energy**

To test each potential's prediction of the total free energy, we use a modified version of the dataset of Zhang, et al.(14). The dataset matches experimental free energies of binding, ΔG, to known crystal structures. We remove from the dataset structures that were determined by NMR, structures that include uracil or an unnatural base, or structures in which there is only one DNA chain in the asymmetric unit. The set then contains the following 30 structures: 1aay, 1apl, 1az0, 1azp, 1bc7, 1bhm, 1bp7, 1ca5, 1cdw, 1cma, 1cw0, 1ecr, 1efa, 1glu, 1hcq, 1hcr, 1ihf, 1ipp, 1lmb, 1mdy, 1nfk, 1oct, 1par, 1pue, 1qrv, 1run, 1tro, 1tsr, 1ysa, and 1ytf. Prediction accuracy is quantified by the correlation between theory and experiment. Because missing side-chains in the structure do not allow an energy calculation on the structure in AMBER, 13 structures were removed for the AMBER calculations (1apl, 1az0, 1bhm, 1efa, 1hcq, 1ihf, 1ipp, 1nfk, 1oct, 1par, 1qrv, 1tro, and 1tsr).

**C. Prediction of changes in free energy**

We use the dataset provided by Morozov, et al.(1) to test the prediction of changes in free energy upon DNA mutation, ΔΔG, in systems with known crystal structures. Several mutants in the set were removed to reflect the size of the binding site in the known crystal structures. The full set includes 189 mutants from ten structures: 1aay, 1ckq, 1ecr, 1efa, 1hcq, 1jk1, 1lmb, 1run, 1tro, and 6cro. 21 mutants from three structures (1efa, 1hcq, and 1tro) are removed for a second set because of missing side-chains in the original structure. For AMBER, because of the smaller number of sequences, we were able to minimize each sequence separately: the pairwise assumption is not used as it is above for the prediction of protein-DNA specificity.

# Results
## I. Derivation of distance dependence and initial comparison of methods

To compare the different knowledge-based potentials, we first use the specificity test on the training set, the set of protein-DNA complexes used to derive the potentials. We ask, how accurately does a potential predict the DNA binding sequence for the structures that were used to derive the potential? Using the test on the training dataset does not necessarily show its predictive ability, but it does represent how accurately each method recapitulates the input data. In a sense, it represents a test of each potential formulation under ideal conditions. The test also allows us to choose a distance cutoff to use for the predictive tests we consider below.

Figure 2 shows the results for the four knowledge-based potentials we consider (quasichemical, DFIRE, μ, and residue) as well as the nonspecific potential (see Methods). The abscissa represents the maximum distance cutoff used for the calculation. Figure 2a displays the results for the methods that use a single distance bin, and Figure 2b displays the results for methods that use multiple distance bins. The quasichemical potential with multiple distance bins is best able to re-predict the bound sequences. The smallest predicted Hamming distance for the three multiple bin methods (quasichemical, DFIRE, and μ) are, respectively, 0.417, 0.503, and 0.495. For a single distance bin methods (quasichemical, residue, and nonspecific) the smallest predicted Hamming distances are, respectively, 0.553, 0.671, and 0.719.

The figure also shows that the most accurate results are found for potentials that use a distance cutoff in the range of 5 to 8 Å. For the single bin methods, the distance cutoffs with the smallest Hamming distance are 5.2 Å for the quasichemical potential, 7.7 Å for the residue potential, and 11.6 Å for the nonspecific potential. For the multiple bin methods, the best-performing distance cutoffs are 7.0 Å for the quasichemical potential, 6.0 Å for the DFIRE potential, and 7.0 Å for the μ potential.

## II. Predictive tests and comparison to existing methods

We next present three predictive tests of the knowledge-based potentials and compare to the results of an existing method, AMBER. The tests are considered predictive because the tested structures were not used to train the knowledge-based potentials, and therefore each method is being asked to make predictions on structures that they have not seen before. The first metric is the same specificity test used above, but now we use the testing set of complexes that were not used for the potential derivations. The second test compares the predicted protein-DNA interaction energies to experimental ΔG values for 30 complexes, and the third test compares the predicted ΔΔG values of 189 DNA mutants from 10 complexes.

Table I presents the results of the three tests for the knowledge-based potentials and the nonspecific potential. The quasichemical potential with multiple distance bins outperforms the other methods for two tests; it gives lower values for the specificity test and higher correlation for the ΔΔG test. It is also important to notice how well the nonspecific potential performs on the ΔG test. Since the predicted binding energy from such a potential scales simply with (and essentially reflects) the size of the interface, the use of the ΔG test as a metric of testing potentials is strongly questionable.

Table II presents the results from the AMBER99 potential, both before and after minimization, for the three tests. Because with AMBER we only consider structures where all side-chains are represented, the datasets are somewhat smaller for the prediction of ΔG and ΔΔG (see Methods).

While AMBER99 does well for the specificity test before minimization, the results after minimization are not nearly as accurate. Neither before nor after minimization does the AMBER99 potential perform similar or better than the quasichemical potential's correlation in the ΔΔG test.

It is also important to note the results when each of the three distinct intermolecular interaction terms in AMBER are calculated separately on the structures. For the specificity test, the Lennard-Jones component does nearly as well alone as in combination with the other terms. For the ΔΔG test, the charge-charge term used on the unminimized structures does better than the full potential which is dominated by the Lennard-Jones component (data not shown).

**III. Database dependence**

Finally, we also consider the database dependence of the knowledge-based potentials. We consider the results of the specificity and ΔΔG tests when only a subset of the complexes in the training set is used to derive the potential. The results were calculated for 100 different random orderings of the training set and then averaged so that the results are independent of the order that the structures are added to the subset.

Considering the results when using smaller subsets can show how well sampled the training set is. Once there are sufficient statistics, the test metric scores when using subsets of increasing size should reach a level value. At this point, the addition of more structures (providing more protein-DNA contact statistics) would not be expected to change the test score. If, on the other hand, the scores continue to improve as structures are added, one expects that a larger structural database will be needed to give the optimal predictive value for the potential.

As shown in Figure 3, the quasichemical potential appears to be approaching, but perhaps not to have reached, a limiting value. Extrapolating these trends to when more protein-DNA structures are known, the prediction of ΔΔG changes and the correct DNA-binding sequence may show a modest improvement as more protein-DNA crystal structures are determined.

## Discussion

We have systematically compared several different protein-DNA energy models. The results show that the models give very different results. We will first discuss the results of the tests as a whole then consider the implications for the various energy models.

### I. Failure of the ΔG test

When considering the three predictive tests, the correlation to experimental ΔG test stands out as a poor metric for testing the potentials. As shown in Table 1, the simplest potential, the nonspecific potential, is able to perform better than any of the potentials tested here. The nonspecific potential is able to predict the free energy merely by counting the protein-DNA contacts within a certain radius. It may be that this number of contacts correlates with the various contributions to total binding free energy in specific complexes(29). This result emphasizes the need for simple controls, such as the nonspecific potential, that differentiate between trivial and nontrivial properties of contact potentials. While the hypothetical true protein-DNA energy function would measure this correctly, simpler models also successfully describe total free energy and could be used for this purpose. When the predicted free energies of lower affinity DNA sequences are compared for the same system, as in the ΔΔG test, the nonspecific potential does not have any predictive value (Table 1).

The other potentials, the specificity test and the ΔΔG test, both appear to be better metrics. Methods other than the nonspecific potential do have predictive value, and the tests clearly differentiate between the potentials.

For the specificity test, the best methods predict more than 60% of the bases correctly. To further improve knowledge-based predictions, accurate direct readout energy models, such as the quasichemical model, need to be combined with a properly weighted and accurate indirect readout term. Water-mediated hydrogen bonds and correctly modeled protein-DNA dynamics upon mutation will also likely be necessary for the most accurate predictions. This is analogous to the result with molecular mechanics potentials, where others have found improved predictions when crystallographic waters are included(10). Likewise, the orientation of interactions will likely need to be explicitly modeled, as others have recently considered(17,30).

While the ΔΔG test clearly differentiates between potentials, the most accurate result calculated in this work gives a correlation coefficient (r) between experimentally measured and calculated ΔΔG values of only 0.46. Even the ROSETTA method, which used this dataset as its training set, can only provide a correlation of 0.57. Ideally, one would find a high correlation coefficient (i.e., r > 0.8). There are reasons, however, to expect a lower correlation for this test. The experimental mutant free energy data is from a variety of experimental methods carried out under different conditions. Also, because we consider only small DNA mutations, free energy differences are small and experimental noise could be an important factor. Nonetheless, for a potential to be useful for ΔΔG prediction a higher correlation will be necessary.

### II. Comparison of knowledge-based potentials
#### A. Different potential derivations and why they result in different potential accuracy

The important difference between the knowledge-based methods is the way in which the potentials are derived. As shown in Figure 4, the potentials give different predictions, for example, for the interaction energy of a very attractive nitrogen-oxygen heavy-atom contact, the main specific contact between an arginine residues and a guanine base. The quasichemical method predicts a much more favorable energy at short distances, while the µ potential predicts a relatively constant attraction and the DFIRE potential predicts attractive energies at short distances and repulsive energies at moderate distances. For protein-DNA interactions, it appears that differences such as these allow the quasichemical method to outperform the µ and DFIRE methods.

It is interesting to consider the reasons that the µ potential works so well at protein folding but not at protein-DNA interactions. The µ potential was originally derived to maximize the gap between the native state and misfolded protein decoys as prescribed by the general theory of protein folding (31,32) . While the chemical characteristics of both protein folding and protein-DNA interactions have many similarities, the physical characteristics of protein folding, such as the hydrophobic collapse and energetic separation of the native state from other states, is unlikely to be directly transferable to the study of protein-DNA interactions. In contrast to protein folding, our understanding of the general physical theory of protein-DNA interactions is much less advanced.

**B. Protein-DNA interactions depend crucially upon an atomic-level description**

For knowledge-based methods, we considered both a heavy atom and a simple residue representation. The heavy atom representation is much more successful than the residue potential at both the specificity and ∆∆G tests. From this we can conclude that accurate descriptions of protein-DNA interactions almost certainly require atom-level information. Another more complex residue potential(4) includes geometric information and would likely outperform the residue method presented here. Because of the more complex molecular representation used, we could not easily include this method in the current work. The geometric information they use is still is less detailed than the heavy atom representation, but future tests will be needed to compare these methods.

**C. Different distance treatments capture larger scale conformational preferences**

While many parameters are selected when any energetic model is developed, we have systematically studied the dependence the methods have on the atom-atom or residue-residue distance. In previous work, distance cutoffs have generally been chosen arbitrarily or based on experimental measurements of atomic size. We sought to determine what the optimal uniform treatment of distance is and how this choice affects the success of the model.

We find two distance regimes do particularly well (Figure 2). First, a short cutoff, such as 4 or 4.5 Å, is relatively successful, particularly for the single bin quasichemical potential. This observation agrees with previous work that describes the importance of short-range contacts (e.g. (33)). Calculations in this regime likely capture the direct atom-atom interactions, such as hydrogen bonding. It appears that in this regime the potential is uncovering the underlying physical interactions of the atom-atom pairs.

A larger cutoff, such as 7 Å, is even more successful for multiple bin potentials. While it is possible that physical long-range interactions may have been captured by the larger cutoff, we cannot rule out that these helpful, long-range contacts may be due to correlations between atoms that are the result of connectivity. For example, in a G-C pair, the guanine ring nitrogen that is a hydrogen bond donor at the Watson Crick interface typically experiences no direct contacts to the protein, but because it is part of the G-C pair, it is frequently observed in long-distance contact with residues that preferentially contact G-C pairs. As a result, these cutoffs capture in a rough way the conformation of the interacting residues and bases. Neighboring atoms not directly in contact with the other polymer now contribute to the energetic prediction, such that more common interaction geometries are more energetically favorable. While the potential may not be accurately describing the physical energy of individual interactions at these larger distances, it appears that including the interactions is technically useful. If, however, additional distance bins are included beyond 7.0 Å, spurious "contacts" from neighboring residues and bases add noise to the predictions. Clearly a single bin does not appear to be enough for these calculations, but a careful choice of multiple distance bins can greatly improve the success of a knowledge-based method.

### III. AMBER: The importance of the Lennard-Jones term and the result of relaxing a crystal structure

Because we consider direct readout, one might expect that the most important part of the AMBER potential would be the charge-charge interactions that describe hydrogen bonding. Instead, the relatively nonspecific Lennard-Jones energy performs better on the specificity test (Table 2). This is very likely due to the repulsive Lennard-Jones energies for DNA mutants. If the mutant structures are not allowed to minimize, certain interactions are in the very repulsive region of the Lennard-Jones potential. This gives a very large energetic penalty, allowing the potential to reject that DNA mutation and select the correct base found in the x-ray crystal structure. If, on the other hand, minimized structures are compared, neither the Lennard-Jones potential nor the full AMBER potential is very predictive (Table 2). Others have similarly found that relaxing structures in AMBER decreases the accuracy of protein-DNA predictions(9).

For the ΔΔG test, the large repulsive Lennard-Jones energies cause the Lennard-Jones and full AMBER potentials to do poorly before full minimization. Both correlations are rather small (Table 2). Minimization removes these repulsive energies, allowing the full AMBER potential to perform somewhat better (r = 0.234), although still much lower than some other methods. If the Lennard-Jones component was less dominant for the non-minimized structures, the correlation with ΔΔG may have been higher.

Morozov, et al.(1) found that when their molecular-mechanics type potential is used to relax the structures, the method is no longer able to predict the free energy of DNA mutations as accurately (correlation drops from 0.57 to 0.4). The high correlation for the unminimized structure with this method is probably possible because at short distances their Lennard-Jones term switches to a linear repulsive term, whereas AMBER increases as $r^{12}$. This allows other, more predictive terms such as directional hydrogen bonding to not be overwhelmed by a steric clash in ROSETTA. While cross-validation on a subset of the ΔΔG data gave similar optimal weights(1), a new set of weights or additional terms may be needed for different types of tests. For example, the method is outperformed by a contact-based model when predicting experimentally determined sequence logos(1).

Statistical potentials, particularly the multiple bin quasichemical method, are able to do comparatively well at both the specificity and ΔΔG tests. These potentials perform well despite not explicitly including terms such as solvation or electrostatics. This allows the potentials to be much faster than the AMBER potential. Because of the necessity to minimize hydrogen positions and calculate the solvation term, AMBER calculations even without full minimization usually took three orders of magnitude longer than the statistical potential calculations. Given the substantial speed advantage and current higher accuracy of knowledge-based contact potentials, these methods show particular promise for large-scale studies of protein-DNA interactions.

**Acknowledgements**

This work was supported by the NIH and NSF. We would also like to thank Chi Zhang, Song Liu, Qianqian Zhu, and Yaoqi Zhou and Alexandre Morozov, James Havranek, David Baker, and Eric Siggia for making, respectively, their ΔG and ΔΔG data freely available. We are thankful to Peter Kutchukian, Jae Shick Yang, James Havranek, and Brian Dominy for helpful conversations and Lucas Nivon for helpful comments on the manuscript.

# References

1. Morozov, A.V., Havranek, J.J., Baker, D. and Siggia, E.D. (2005) *Nucleic Acids Res*, **33,** 5781-5798.
2. Michael Gromiha, M., Siebers, J.G., Selvaraj, S., Kono, H. and Sarai, A. (2004) *J Mol Biol*, **337,** 285-294.
3. Havranek, J.J., Duarte, C.M. and Baker, D. (2004) *J Mol Biol*, **344,** 59-70.
4. Kono, H. and Sarai, A. (1999) *Proteins*, **35,** 114-131.
5. Mandel-Gutfreund, Y., Baron, A. and Margalit, H. (2001) *Pac Symp Biocomput*, 139-150.
6. Mandel-Gutfreund, Y. and Margalit, H. (1998) *Nucleic Acids Res*, **26,** 2306-2312.
7. Gromiha, M.M. and Selvaraj, S. (2001) *J Mol Biol*, **310,** 27-32.
8. Olson, W.K., Gorin, A.A., Lu, X.J., Hock, L.M. and Zhurkin, V.B. (1998) *Proc Natl Acad Sci U S A*, **95,** 11163-11168.
9. Endres, R.G., Schulthess, T.C. and Wingreen, N.S. (2004) *Proteins*, **57,** 262-268.
10. Endres, R.G. and Wingreen, N.S. (2006) *Phys Rev E Stat Nonlin Soft Matter Phys*, **73,** 061921.
11. Paillard, G., Deremble, C. and Lavery, R. (2004) *Nucleic Acids Res*, **32,** 6673-6682.
12. Arauzo-Bravo, M.J., Fujii, S., Kono, H., Ahmad, S. and Sarai, A. (2005) *J Am Chem Soc*, **127,** 16074-16089.
13. Cheng, A.C. and Frankel, A.D. (2004) *J Am Chem Soc*, **126,** 434-435.
14. Zhang, Y. and Skolnick, J. (2005) *Proc Natl Acad Sci U S A*, **102,** 1029-1034.
15. Kussell, E., Shimada, J. and Shakhnovich, E.I. (2002) *Proc Natl Acad Sci U S A*, **99,** 5343-5348.
16. Lu, H. and Skolnick, J. (2001) *Proteins*, **44,** 223-232.
17. Miyazawa, S. and Jernigan, R.L. (1996) *J Mol Biol*, **256,** 623-644.
18. Zhou, H. and Zhou, Y. (2002) *Protein Sci*, **11,** 2714-2726.
19. Zhou, Y., Zhou, H., Zhang, C. and Liu, S. (2006) *Cell Biochem Biophys*, **46,** 165-174.
20. Chen, W.W. and Shakhnovich, E.I. (2005) *Protein Sci*, **14,** 1741-1752.
21. Hubner, I.A., Deeds, E.J. and Shakhnovich, E.I. (2005) *Proc Natl Acad Sci U S A*, **102,** 18914-18919.
22. Berman, H.M., Westbrook, J., Feng, Z., Gilliland, G., Bhat, T.N., Weissig, H., Shindyalov, I.N. and Bourne, P.E. (2000) *Nucleic Acids Res*, **28,** 235-242.
23. Wang, J.M., Cieplak, P. and Kollman, P.A. (2000) *Journal of Computational Chemistry*, **21,** 1049-1074.
24. Ren, P. and Ponder, J.W. (2003) *Journal of Physical Chemistry B*, **107,** 5933-5947.
25. Ponder, J.W. and Richards, F.M. (1987) *Journal of Computational Chemistry*, **8,** 1016-1024.
26. Qui, D., Shenkin, P.S., Hollinger, F.P. and Still, W.C. (1997) *J. Phys. Chem. A.*, **101,** 3005-3014.
27. Lu, X.J. and Olson, W.K. (2003) *Nucleic Acids Res*, **31,** 5108-5121.
28. Qian, J., Luscombe, N.M. and Gerstein, M. (2001) *J Mol Biol*, **313,** 673-681.
29. Ishchenko, A.V. and Shakhnovich, E.I. (2002) *J Med Chem*, **45,** 2770-2780.

30. Buchete, N.V., Straub, J.E. and Thirumalai, D. (2004) *Curr Opin Struct Biol*, **14,** 225-232.
31. Shakhnovich, E. (2006) *Chem Rev*, **106,** 1559-1588.
32. Ramanathan, S. and Shakhnovich, E. (1994) *Physical Review. E. Statistical Physics, Plasmas, Fluids, and Related Interdisciplinary Topics*, **50,** 1303-1312.
33. Li, X., Hu, C. and Liang, J. (2003) *Proteins*, **53,** 792-805.

**Figures legends**

Figure 1. Schematic showing the results of a single base pair mutation in our structural representation. A T-A base pair is mutated to the other three possible mutants (G-C, C-G, and A-T). The original structure has the PDB code 1a02.

Figure 2. Distance dependence of the specificity test on the training set for the knowledge-based potentials. The specificity test measures the number of mismatches between the DNA sequence with the lowest predicted energy and the sequence used in the crystallized structure. The number of mismatches is referred to as the Hamming distance. The distance is normalized by the chain length and averaged over the structures. Because the normalized Hamming distance represents the number of incorrectly predicted bases, smaller values show better predictions. (A) Results for potentials that use a single distance bin. (B) Results for potentials that use multiple distance bins.

Figure 3. Database dependence of the quasichemical potential with multiple distance bins (0-7 Å). The dashed line represents the final value when the full training set is used. (A) Results for the specificity test on the testing set. Smaller values represent more accurate predictions. (B) Results for the $\Delta\Delta G$ test. The $\Delta\Delta G$ test measures the correlation between predicted and experimentally determined free energy changes when DNA bases are mutated. Because the $\Delta\Delta G$ test measures the correlation with experiment, larger values show better predictions.

Figure 4. Predicted energies of a very attractive protein-DNA atomic interaction, a guanine oxygen with an arginine nitrogen, in different distance bins for the three heavy-atom knowledge-based potentials (quasichemical, $\mu$, and DFIRE).

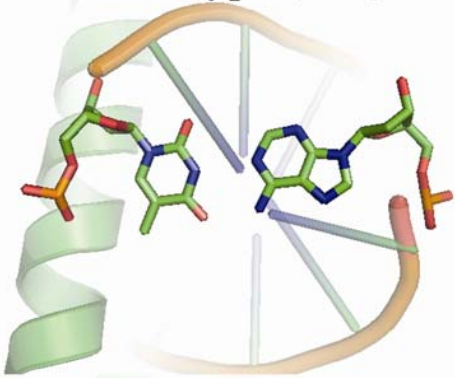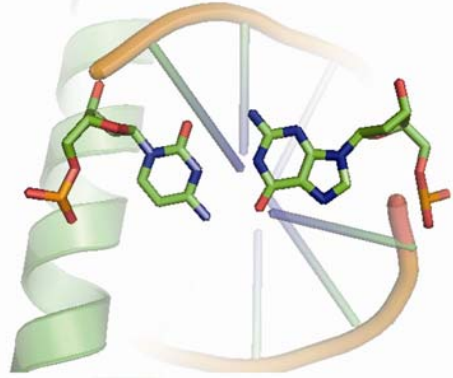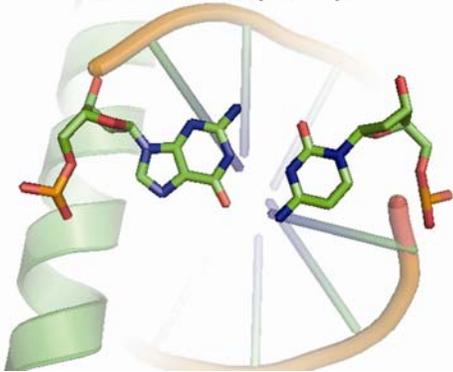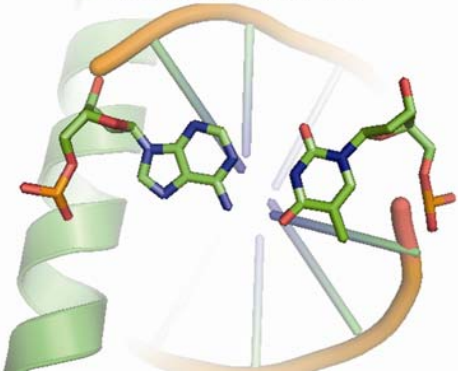

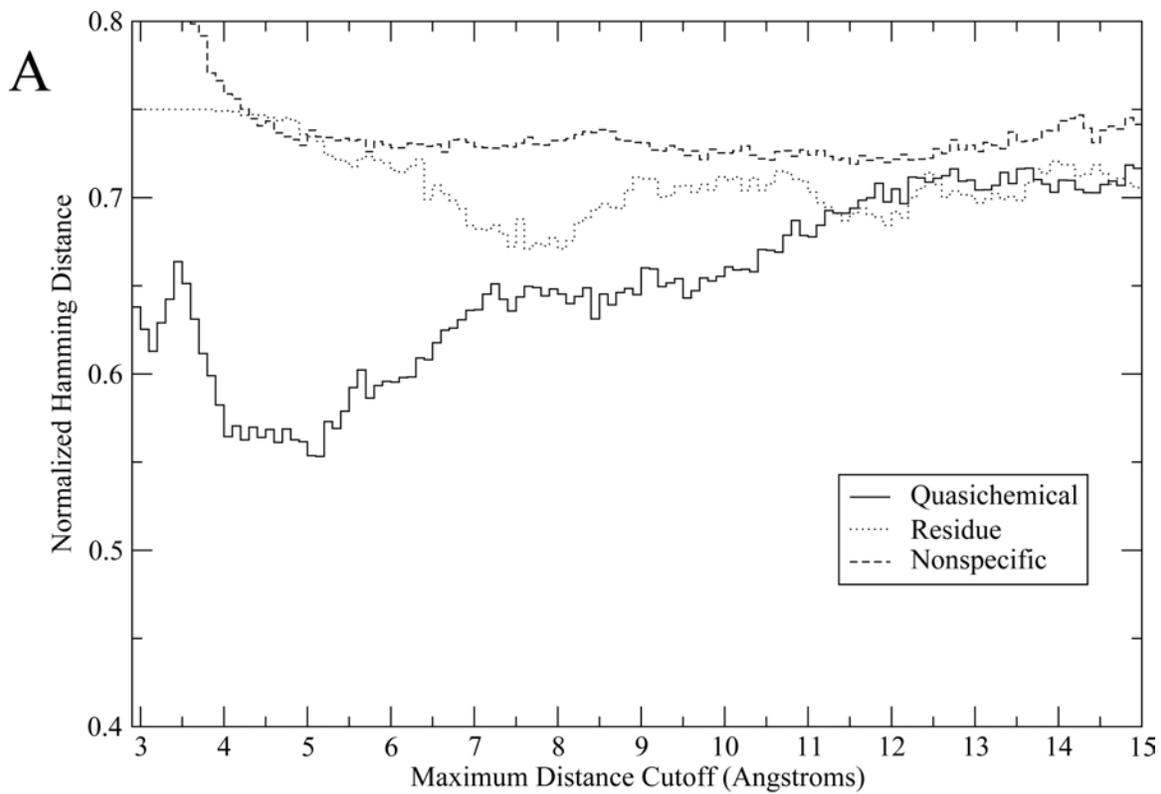

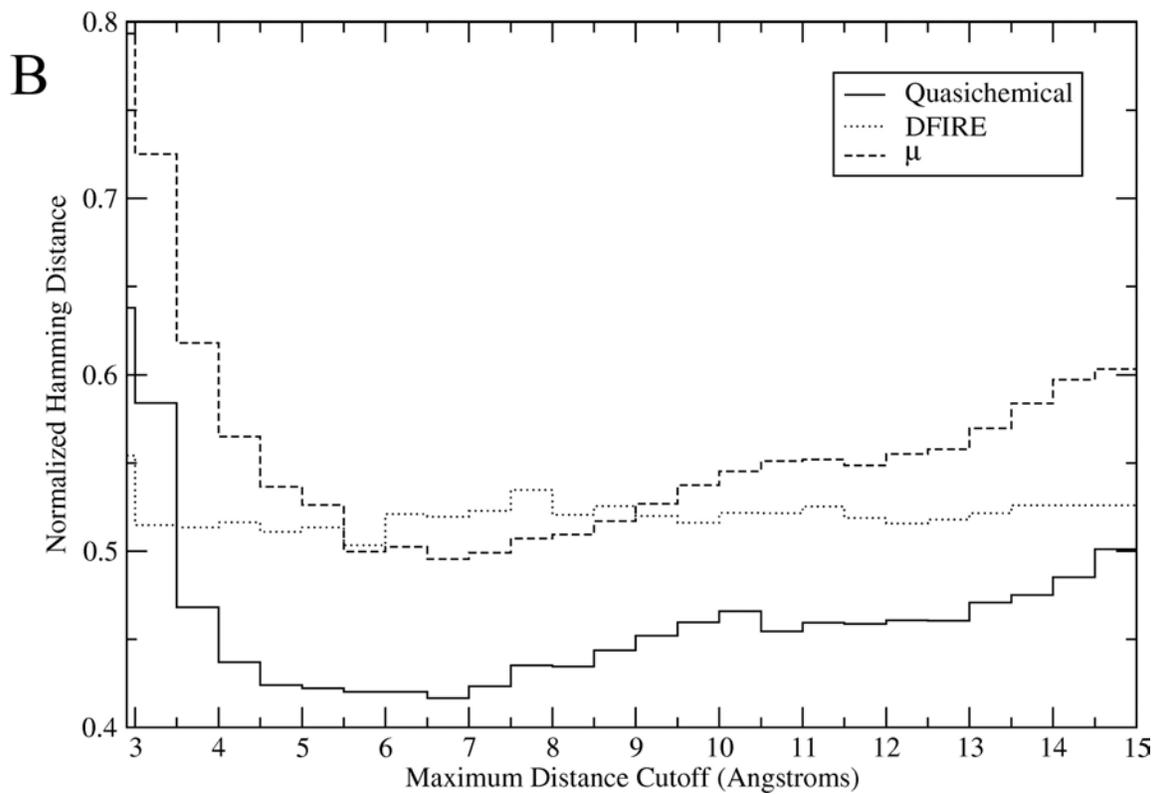

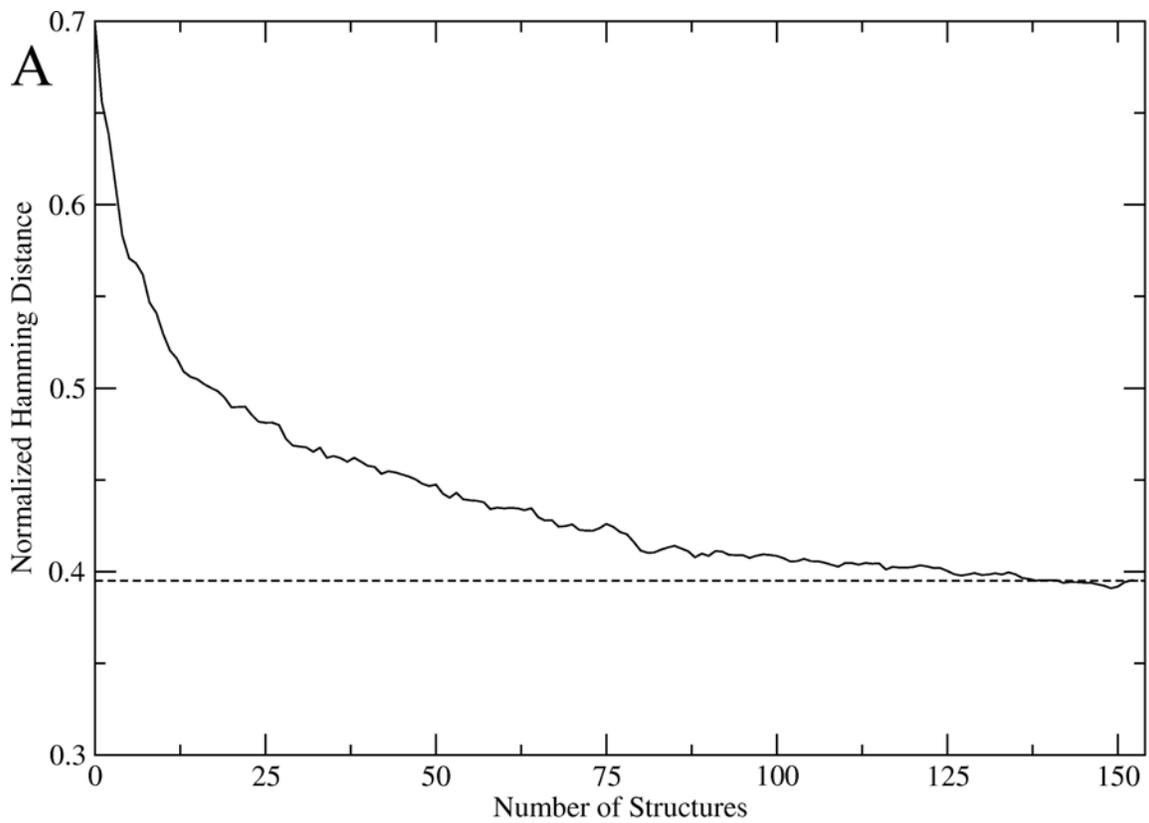
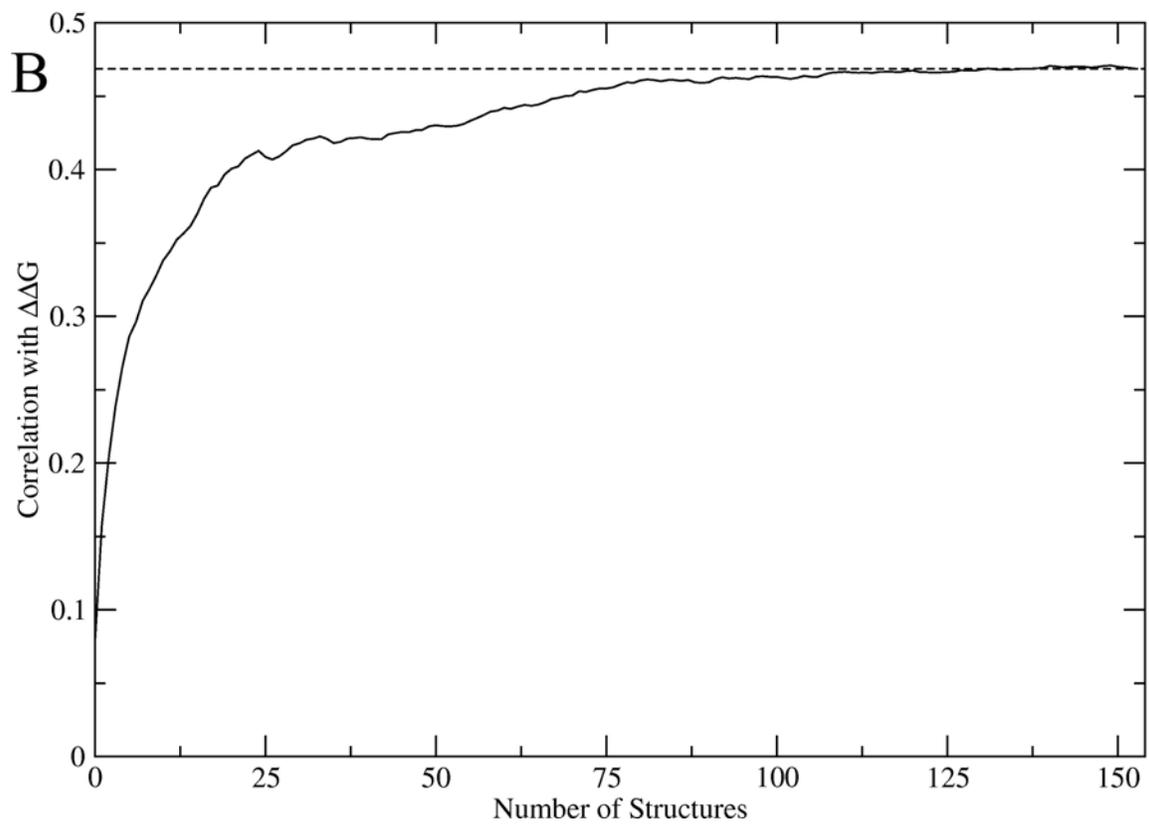

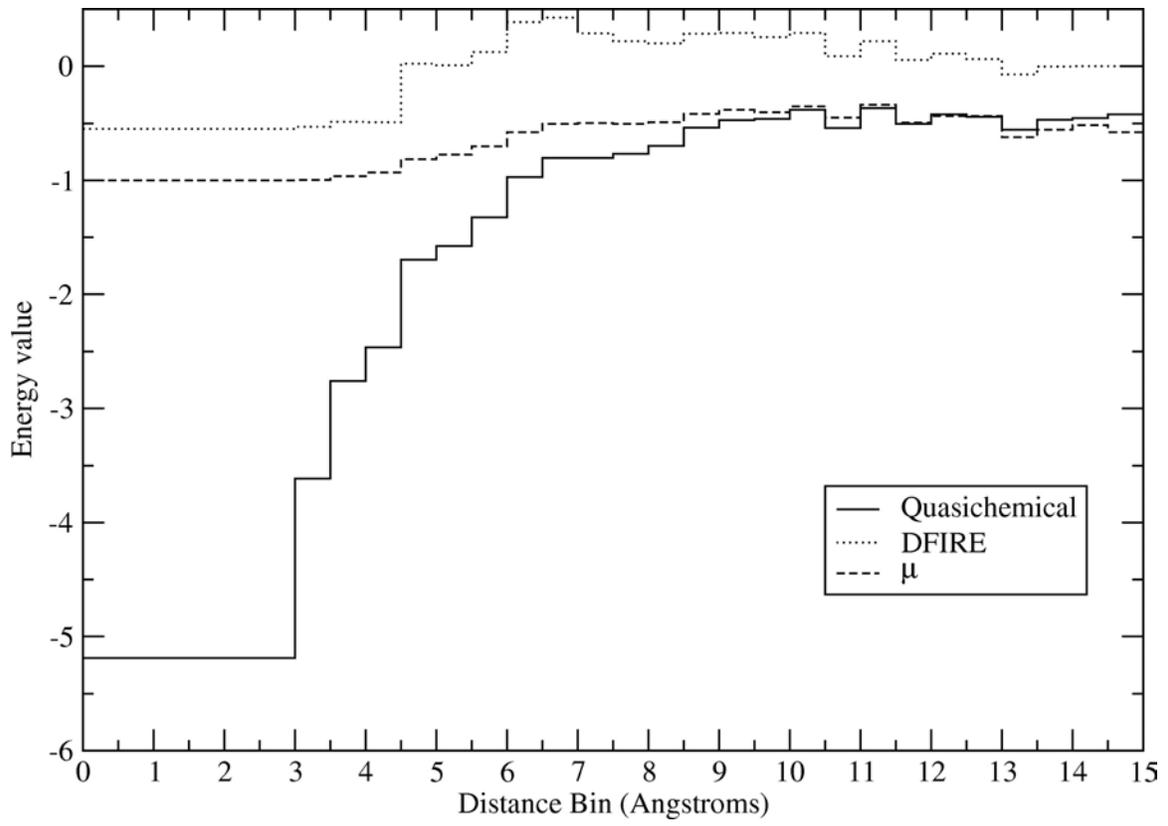

**Tables**

|  | Specificity test | ΔG test | Δ ΔG test |
|---|---|---|---|
| Single bin | | | |
| Quasichemical | 0.494 | 0.265 | 0.117 |
| Residue | 0.720 | 0.331 | -0.078 |
| Nonspecific | 0.723 | 0.829 | -0.116 |
| Multiple bin | | | |
| Quasichemical | 0.395 | 0.233 | 0.468 |
| DFIRE | 0.530 | -0.555 | 0.364 |
| μ | 0.461 | 0.526 | 0.317 |
| For reference | | | |
| Ideal | 0 | 1 | 1 |
| Random | 0.75 | 0 | 0 |

Table 1. Results of the three predictive tests. The specificity test measures the number of mismatches between the predicted optimal DNA sequence and that found in the crystal structure, normalized by the sequence length and averaged over the structures. The ΔG and ΔΔG tests consider the correlation between the predicted free energy or free energy change and that observed experimentally. For reference, both an "ideal" result and the expected result of a random method are listed.

|  | Specificity test | ΔG test | Δ ΔG test |
|---|---|---|---|
| AMBER |  |  |  |
| **Before full minimization** | **0.330** | **0.778** | **0.141** |
| *Lennard-Jones* | *0.349* | *0.848* | *0.141* |
| *Charge-charge* | *0.558* | *0.547* | *0.432* |
| *Solvation* | *0.823* | *-0.553* | *-0.359* |
| **After full minimization** | **0.670** | **0.757** | **0.234** |
| *Lennard-Jones* | *0.627* | *0.845* | *0.037* |
| *Charge-charge* | *0.658* | *0.525* | *-0.024* |
| *Solvation* | *0.879* | *-0.530* | *0.048* |
| Knowledge-based |  |  |  |
| Quasichemical | 0.395 | 0.125 | 0.462 |
| For reference |  |  |  |
| Ideal | 0 | 1 | 1 |
| Random | 0.75 | 0 | 0 |

Table 2. Results of the three predictive tests for AMBER and the multiple-distance bin quasichemical potential. Because of missing side-chains in some of the structures in the original dataset, the ΔG and ΔΔG tests use smaller datasets (see Methods). For AMBER, the results for each of the three components of intermolecular interactions are also presented in italics while the results for the full potential are in bold.